\documentclass[traditabstract]{aa} 
\usepackage{graphicx}
\usepackage{txfonts}
\usepackage{epsfig}                            
\usepackage{natbib}

\begin{document}
%
   \title{Transit mapping of a starspot on CoRoT-2}

   \subtitle{Probing a stellar surface by planetary transits}

   \author{U. Wolter \inst{1},
          J.H.M.M. Schmitt \inst{1},
          K.F. Huber \inst{1},
          S. Czesla \inst{1},
          H.M. M\"uller \inst{1},
          E.W. Guenther \inst{2}
          \and
          A.P. Hatzes \inst{2}          
          }
   \authorrunning{Wolter, Schmitt et al.}

   \offprints{U. Wolter, \\ \email{uwolter@hs.uni-hamburg.de}}

   \institute{Hamburger Sternwarte, Gojenbergsweg 112, D-21029 Hamburg, Germany \\
              \email{uwolter@hs.uni-hamburg.de, jschmitt@hs.uni-hamburg.de}
       \and
              Th\"uringer Landessternwarte Tautenburg, Sternwarte 5, D-07778 Tautenburg, Germany\\
              \email{guenther@tls-tautenburg.de, artie@tls-tautenburg.de}
             }

   \date{Received ... / Accepted ...}  

   \abstract{
We analyze variations in the transit lightcurves of CoRoT-2b,
a massive hot Jupiter orbiting a highly active G~star.
We use one transit lightcurve to eclipse-map a photospheric spot
occulted by the planet.
In this case study we determine
the size and longitude of the eclipsed portion of the starspot
and systematically study the corresponding uncertainties.
We determine a spot radius between $4.5\degr$ and~$10.5\degr$
on the stellar surface and the spot longitude
with a precision of about $\pm 1$~degree. 
Given the well-known transit geometry of the CoRoT-2 system, this implies
a reliable detection of spots on latitudes typically covered by sunspots;
also regarding its size the modelled spot is comparable
to large spot groups on the Sun.
We discuss the future potential of eclipse mapping by planetary transits for 
the high-resolution analysis of stellar surface features.
%
   }


   \maketitle
%

\section{Introduction}
\vspace*{-1.0\medskipamount}
The atmosphere of the Sun shows inhomogeneities down to the smallest scales 
currently accessible to solar observations which are of the order of 50~km 
(e.g. \citealt{Scharmer02}).
Such intricate fine structure can also be
expected in the atmospheres of other active stars.
However, the best currently available stellar observations 
only resolve surface features down to the size of a few degrees on the
surface, corresponding to several 10000~km on a main sequence star.

The increasing number of known transiting extrasolar planets 
offers 
an outstanding
opportunity to
study surface inhomogeneities 
\textit{of their host stars} with an unprecedented surface resolution. 
As an example of this new technique, we present the analysis of one 
transit lightcurve
of the planetary system CoRoT-2, 
recently detected by the CoRoT satellite \citep{Rouan98}.
Our study indicates that under favourable conditions the CoRoT lightcurves
allow the study of e.g. starspots and/or
faculae down to a sub-degree scale on the stellar surface. 

Deformations of planetary transit lightcurves, attributable to dark spots,
have been observed for several systems: 
HD~189733, \citealt{Pont07}; 
HD~209458, \citealt{Silva08};
Tres-1,   \citealt{Rabus09}   
and CoRoT-2, \citealt{Lanza09}.
Especially, 
\citeauthor{Pont07}'s study, based on HST data, for the first time showed that
low-noise transit photometry of exoplanets can yield detailed information
about surface features of their host star.
We present a 
systematic analysis of the spot 
locations and extensions, including their uncertainties,
that can be deduced from transit
lightcurves.


\smallskip
CoRoT-2a \mbox{(GSC 00465-01282)} is an apparently young late  G-dwarf star (\citealt{Alonso08}, AL08 in the following);
it is unsually active and intrinsically variable among the 
presently known planetary host stars.
Its rotation period of \mbox{$P_{\ast}=4.52\pm0.14~d$} (\citealt{Lanza09}) 
amounts to less than three times the planetary orbit period $P_{orb}$:
\mbox{$13\cdot P_{orb} = 5.004 \cdot P_{\ast}$}.
Densely sampled high-precision photometry in combination with 
spectroscopic measurements 
make the transit geometry of 
CoRoT-2b exceptionally well known (AL08, \citealt{Bouchy08}):
The planet is on a nearly circular orbit (\mbox{$e=0.003\pm0.003$}) 
which is seen close to edge-on
(\mbox{$i=87.84\pm0.10\degr$}) and approximately oriented perpendicular to the 
stellar rotation axis (deviating by \mbox{$\lambda=7.2\pm4.5\degr$}).

These attributes turn the planetary disk into an \textit{extremely well-defined  
probe} which periodically scans 
a band on the stellar surface covering
$20\degr$ in latitude.
The orbital period of CoRoT-2b translates into 
an orbital angular velocity of 0.002~deg/s.
As a result, close to the center of the stellar disk,
the planet moves \mbox{$\approx 0.1$~deg} in longitude
 during a 32~s time bin of the CoRoT data.
By order of magnitude, this value then determines the attainable 
surface resolution of the CoRoT observations of CoRoT-2.

\begin{figure*} [ht!]
\center{
  \epsfig{file=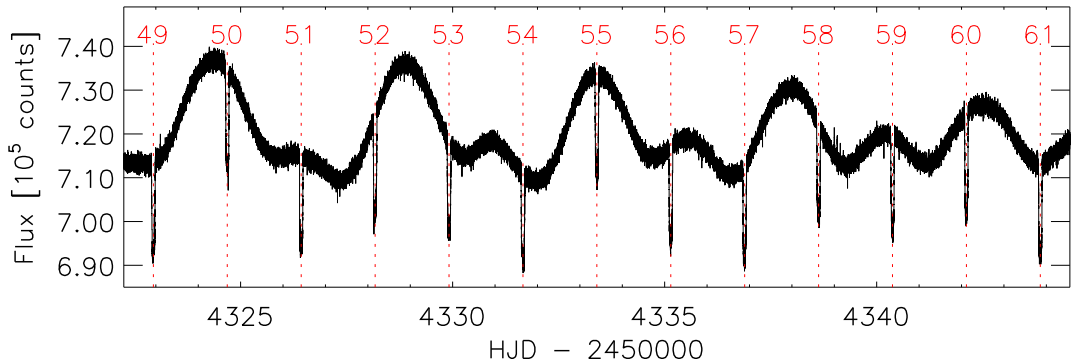, width=0.80\linewidth,clip=, bb=10 10 445 153}
\vspace*{-0.1cm}
}
\caption{
Section of the CoRoT lightcurve of CoRoT-2,
purged of outlying values (see Sect.~\ref{sec:Obs+Analysis}).
The transits appear as recurrent dips 
overlying the modulation by the stellar rotation.
Transits are
numbered from the beginning of the CoRoT observations,
our analysis focusses on transit no.~56.
       }
\label{fig:LCoverview}
\end{figure*}

\section{Observations and data analysis}
\label{sec:Obs+Analysis}
The CoRoT \citep{Auvergne09}
lightcurve of CoRoT-2 observed from
2007~May~16 to 2007~October~15 
continuously samples 31 stellar rotations.
After five days of observations the satellite's sampling cadence 
of CoRoT-2 was switched from 512~s to
32~s,
resulting in a lightcurve 
that covers 79 planetary transits at this high time resolution.

Our analysis started out from the lightcurve data as delivered by the CoRoT pipeline.
The pipeline sorts out defective data
(e.g. due to crossings of the South Atlantic Anomaly, SAA) 
and performs a background subtraction.
As a first step we added up all three CoRoT photometry channels 
(red-green-blue) into a single lightcurve,
because the individual channels are more
affected by instrumental instabilities than their summed-up signal.

Similar to AL08, we then removed outlier points 
which show a pronounced non-normal distribution.
To this end we 
computed
the standard deviation~$\sigma$ in several 
narrow intervals,
yielding \mbox{$\sigma \approx 950$}.
We discarded
all points deviating more than 
$3\sigma$ from a boxcar-smoothed copy of the lightcurve,
thus rejecting nearly 2\% of all points. 
Finally, also following~AL08, we corrected for the $5.6\%$~contamination by a neighbouring 
object;
a subinterval
of the resulting lightcurve
is shown in Fig.~\ref{fig:LCoverview}.

\begin{figure}
\center{
 \epsfig{file=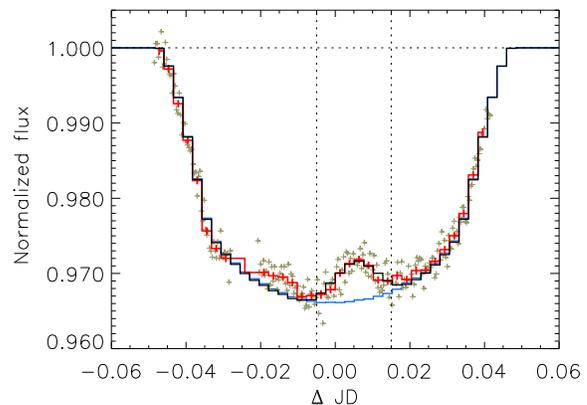, width=0.90\linewidth,clip=}   
}
\vspace*{-1.0\medskipamount}
\caption{
Normalized lightcurve 
during ``transit~56'', 
as a function of time from the transit center.
Gray symbols indicate the 
unbinned measurements,
the red line shows them averaged in 224~sec time bins
with 1$\sigma$-errors indicated.
The vertical lines delimit the time interval used for our spot model;
the lightcurve fit resulting from our model 
is drawn black 
(model ``BN'' in Table~\ref{tab:SpotSolutions}).
The blue line shows the transit model 
for an unspotted star for comparison.
The 
shallow deformation left of the transit center 
is caused by another spot 
not included in our model.
        }
\label{fig:TransitLC}
\end{figure}

The transits occur during different stellar rotation phases, i.e. at
different levels and slopes of the lightcurve.
In order to compare the transit lightcurves, 
it is convenient to normalize them
to a common level. To this end, we
carried out a linear interpolation of 
points adjacent to each transit and divided all transit 
lightcurves by their interpolating function.
One lightcurve normalized in this way is shown 
in Fig.~\ref{fig:TransitLC}.  

We note that this procedure introduces a systematic error that depends on the
spot coverage of the stellar disk visible during the transit.
If, e.g., the disk is covered by dark spots not occulted by the planet,
the transit depth will not yield the true ratio of radii $R_{planet}/R_{\ast}$.
Instead, the stellar radius $R_{\ast}$ will be underestimated relative to the planet, because
part of the stellar disk is dark and essentially invisible.
Given the typical amplitude of CoRoT-2's lightcurve of about 4\% peak-to-peak,
this introduces a comparable uncertainty for the planetary radius
deduced from a single transit.
Consequently, this uncertainty also affects our spot size estimate 
of the next section.
We will discuss the influence of activity-induced lightcurve variations
on the determination of planetary parameters in a forthcoming paper.

\section{Transit lightcurve modelling}
\label{sec:LCModelling}
\vspace*{-0.85\medskipamount}

To determine a spot configuration
compatible with the deformations of a transit lightcurve,
we selected one particular transit
occuring close to JD~2454335.0 and referred to as ``transit~56''.
Rendered in Fig.~\ref{fig:TransitLC}, it
shows the most pronounced and isolated ``bump'' of the whole
time series of CoRoT-2, suggesting a relatively narrow spot
occulted close to the disk center.
Additionally, the symmetry of the bump indicates a
spot geometry that is largely symmetric in longitude.

\subsection{The model}
\label{sec:Model}
\vspace*{-0.85\medskipamount}
To model a transit lightcurve we decompose the stellar surface
into roughly square elements.
The integrated stellar flux is computed by summing up the
fluxes of all visible elements, respecting
their projected area 
and the limb darkening (e.g. \citealt{Wolter05}). 
To model the planetary transit, all surface elements 
occulted by the planetary disk at a given phase are removed
from the sum,
thus treating the planet as a circular disk without any intrinsic emission.
The surface resolution needs to be sufficient to 
analyze the densely sampled movement of the planetary disk over the
stellar surface. 
We choose a surface grid with 750 elements at the equator, 
yielding 178\,868 elements in total   
and a surface resolution of roughly $0.5\degr$ near the
equator. Also, this high resolution is required to compute lightcurve 
models sufficiently free of jitter,
not exceeding a few $10^{-4}$ in our models.

Next, we adjusted the limb darkening to optimize the fit
to the transit 
ingress and egress 
as well as to a tentative lower envelope of
transit~56 and the surrounding transits.
We adopt a linear limb-darkening law with $\epsilon=0.6$;
we note, however, that this has little influence on our determined spot parameters
since the spot occultation in transit~56 occurs close to the disk center.

We compute the position of the planetary disk 
on the stellar disk and the corresponding stellar rotation phase
using the 
orbital parameters and planetary size given by AL08, as well as
a stellar rotation period of \mbox{$P_\ast = 4.5284$~d},
determined by a periodogram analysis and
compatible with the value of \citet{Lanza09},
adopting the lightcurve maximum at JD~2454242.67 as phase zero point.
To this end, 
we introduce Cartesian coordinates in units of the stellar radius $R_\ast$
with their origin at the center of the star and 
the observer located on the $x$-axis. 
The stellar rotation axis is assumed to 
coincide with the $z$-axis, while the axis of the planetary orbit is 
tilted by  \mbox{$90\degr - i = 2.2\degr$} in the $xz$-plane.
Both, the stellar rotation and the planetary orbit are adopted as right-handed
around the $z$-axis.

Using these coordinates,  $z_{pl}$ and $y_{pl}$
describe the position of the center of the planetary disk projected onto the
stellar disk visible at the stellar rotation phase~$\phi$.
The following relations apply:

\begin{equation}
  \hspace*{5 ex} z_{pl} = \cos{i \cdot a / R_{\ast}}  \\
\label{eq:planet_z}
\end{equation}
where $i$ and $a$ describe the inclination and the semimajor axis of
the planetary orbit, respectively.

%
%
\begin{equation}
  \hspace*{5 ex} \phi - \phi_{cen} = \frac{y_{pl}}{y_{IV}} \cdot \frac{\tau_{trans}}{P_{\ast}}
  \label{eq:transit_phase}
\end{equation}
where $\phi_{cen}$ is the stellar rotation phase at transit center, while $\tau_{trans}$ 
and $P_{\ast}$ give the transit duration (first to last contact) and the stellar
rotation period, respectively.
\mbox{$\tau_{trans} = 8200$~s} can be computed using Eq.~(4) of \cite{Charbonneau06}.
$y_{IV}$ describes the lateral planet position at last contact
\begin{equation}
  \hspace*{5 ex} y_{IV} = \sqrt{\left ( 1 + R_{pl}/R_{\ast} \right )^2 - z_{pl}^2}
\end{equation}
with $R_{pl}$ giving the planetary radius.
Note that Eq.~\ref{eq:transit_phase} is only strictly valid for
\mbox{$R_{\ast} \ll a$}; however, the error is of the order of
\mbox{$(R_{\ast} / a)^2$} and can be neglected for our analysis.
In addition, Eq.~\ref{eq:planet_z} does not include 
the projected angle between the stellar rotation axis
and the planetary orbital axis $\lambda$. However, given 
\citet{Bouchy08}'s value of $\lambda \approx 7\degr$ and the proximity of
the modelled spot to the disk center, this introduces an uncertainty of
less than $2\degr$ for the spot latitude.

%

\begin{figure}
\center{
 \epsfig{file=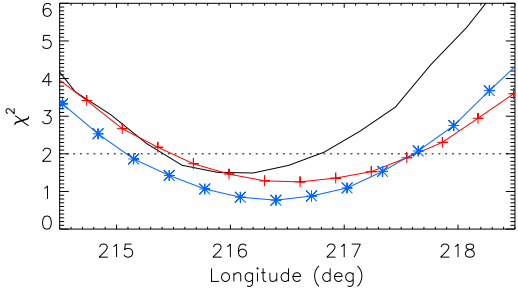, width=0.90\linewidth,clip=}  
}
\vspace*{-1.0\medskipamount}
\caption{
Fit quality of our transit lightcurve models
as a function of spot longitude,
shown for the models BC (blue stars), BN (red plus symbols) and DS (black curve),
of Table~\ref{tab:SpotSolutions}.
Our adopted limiting value of  
\mbox{$\chi^2=2.0$}
is marked by the dotted line.
        }
\label{fig:X2longs}
\end{figure}

\begin{figure}
\center{
 \epsfig{file=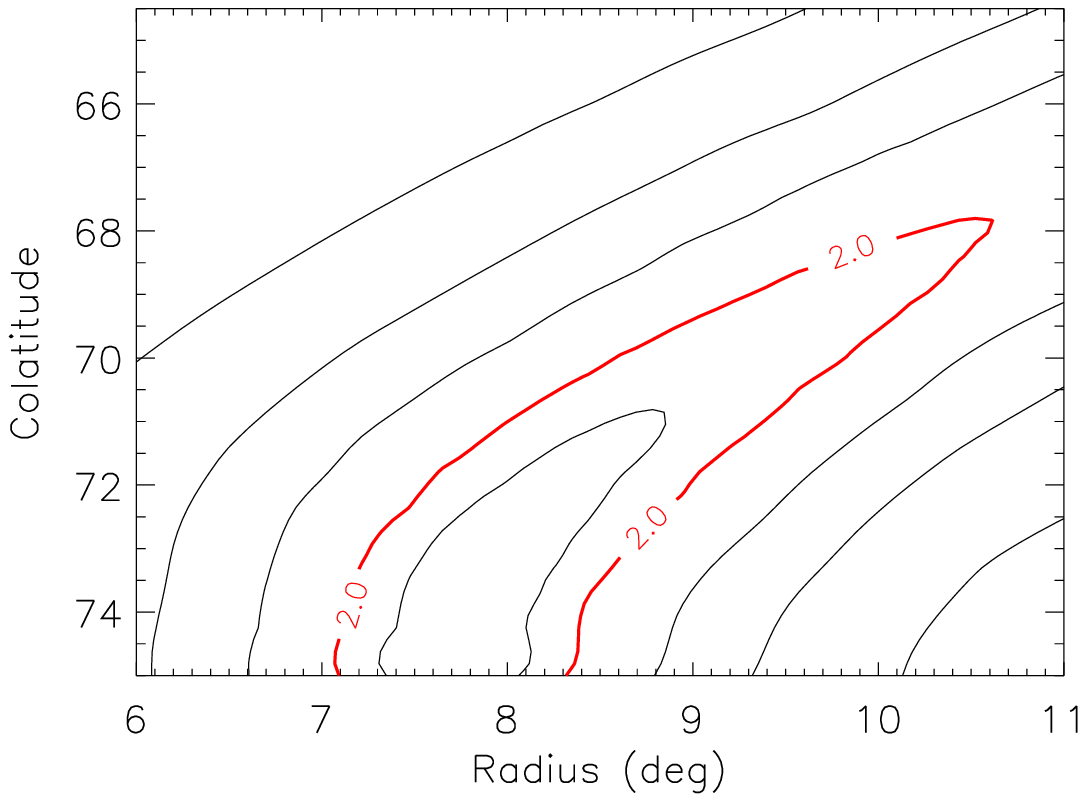, width=0.85\linewidth,clip=, bb=15 25 345 245} 
 \epsfig{file=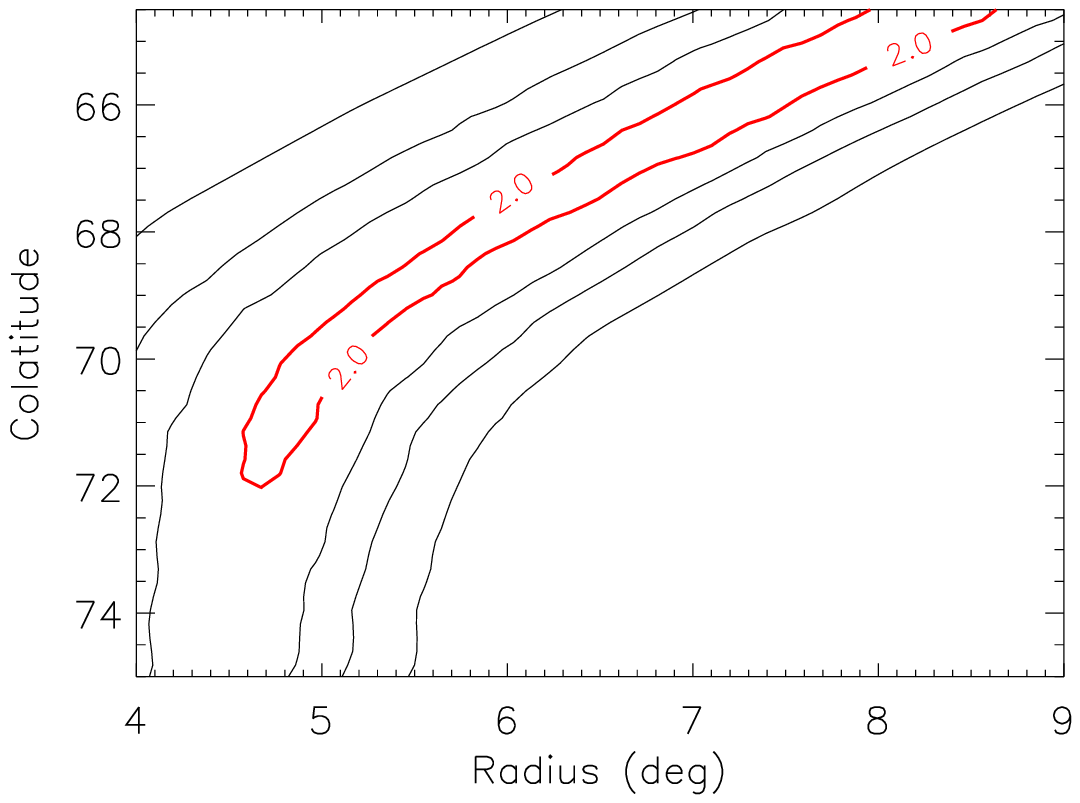, width=0.85\linewidth,clip=, bb=15 0 345 245} 
}
\vspace*{-1.0\medskipamount}
\caption{
Fit quality~$\chi^2$ of our transit lightcurve models
as a function of the spot radius and its center colatitude. 
The \textit{upper panel} describes the ``bright spot'' scenario
(models BC, BN and BS of Table~\ref{tab:SpotSolutions}),
while the \textit{lower panel} applies to the ``dark spot'' solutions
(models DN and DS).
Subsequent contours mark levels of \mbox{$\chi^2=1, 2, 5, 10$} and~$20$;
the adopted limiting contour, \mbox{$\chi^2=2$}, is drawn bold red.
Because of the near symmetry of the $\chi^2$-contours, only colatitudes
in the upper half of the transit band are shown.
See text for discussion.
        }
\label{fig:X2contours}
\end{figure}

\subsection{Transit mapping of a starspot}
\label{sec:EclipseMap}
\vspace*{-0.85\medskipamount}
Our initial tests showed that 
the given transit lightcurves do 
not significantly constrain the spot shape and
contrast.
Hence we limit our models to circular spots
or circle segments.
Furthermore, 
we tentatively adopt two values for the spot-to-photosphere contrast:
As a ``dark spot'' scenario we choose a spot flux of 30\% of the
photospheric flux. 
Based on \citet{Bouchy08}'s effective temperature of CoRoT-2a of $5625$~K and Planck's
law at a wavelength of $6000$~\AA, 
this corresponds to a spot approximately $1200$~K cooler than the photosphere.
This is motivated by spot temperatures found for other
highly active G and K~stars (e.g. \citealt{Strassmeier98}, \citealt{ONeal04}).
As a ``bright spot'' we adopt a value of 75\% of the
photospheric flux. 
This roughly represents the average contrast of large spot groups 
on the Sun at visible wavelengths (\citealt{Walton03}, \citealt{Albregtsen84}).
Our two spot contrast scenarios are comparable to those adopted for CoRoT-2a
by \citet{Lanza09}.


Defining the spot contrast and shape in this way, three
free parameters describe a given spot, namely the spot radius~$r$ as well as
its central longitude~$\varphi$ and colatitude~$\theta$.
However, as illustrated by Fig.~\ref{fig:X2longs},
the longitude is closely confined by the transit lightcurve, 
irrespective of the spot contrast and colatitude.
This is due to the well defined maximum of the transit bump.
Thus only two undetermined parameters remain to optimize 
the fit of transit~56's lightcurve: $r$ and~$\theta$.
We calculated model lightcurves for grids in these two parameters,
the resulting goodness-of-fit $\chi^2$ is shown in Fig.~\ref{fig:X2contours}.

To calculate $\chi^2$ we rebinned the lightcurve
into 224~sec bins, and estimated the resulting errors $\sigma_{j}$
assuming Gaussian error propagation:
\mbox{
$
\chi^2 = {1}/{(N - M)} \cdot
                            {\sum_{j=1}^{N} \left(F_{obs,j} - F_{model,j}\right)^2 } /
                                  {\sigma_{j}^2}
$
}.
Here, $M=3$ is the number of free model parameters, 
namely the spot longitude, colatitude and radius.
$N=8$ gives the number of time bins used to constrain the modelled spot, 
see the interval limits indicated in Fig.~\ref{fig:TransitLC}.

Reducing $\chi^2$ in this way is non-unique because of the  
significant correlation of the spot radius and colatitude,
illustrated by the slanted contours of Fig.~\ref{fig:X2contours}.
Tentatively we choose a limit of \mbox{$\chi^2 \le 2.0$}
for models judged as compatible with the observations.
As an example, Fig.~\ref{fig:TransitLC} shows a solution yielding
\mbox{$\chi^2=1.4$}.

\subsection{Results}
\label{sec:Results}
\begin{figure} [t!]
\center{
 \epsfig{file=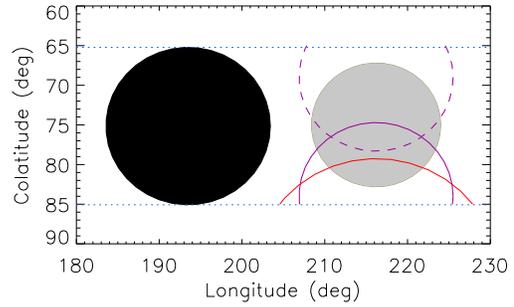, width=0.75\linewidth,clip=, bb=5 110 355 330} 
}
\caption{
A starspot on CoRoT-2a occulted by the planet during transit~56,
as reconstructed by transit mapping.
The black and
gray circles represent the planetary disk and the spot
for our ``bright spot'' scenario, i.e. adopting a 
spot flux of 75\% relative to the photosphere.
The purple arcs illustrate the northern- and  
southernmost solutions for this spot flux.
The red arc illustrates a ``dark spot'' solution 
(DEQ in Table~\ref{tab:SpotSolutions}),
see text for discussion.
        }
\label{fig:TransitMap}
\end{figure}

\vspace*{-0.85\medskipamount}
As the $\chi^2$-contours in the upper panel of Fig.~\ref{fig:X2contours} show,
for the ``bright spot'' scenario, the central spot colatitude 
is confined to~\mbox{$\theta = 75\pm6\degr$}, i.e. inside the stellar surface belt transited
by the planet.
The spot radii are slightly smaller than, or comparable
to, the size of the planetary disk:
\mbox{$r = 7.1\degr \ldots 10.6\degr$}.
This scenario is illustrated by Fig.~\ref{fig:TransitMap} and
the exemplary solutions BN, BC and BS 
(``bright north, central and south '') of Table~\ref{tab:SpotSolutions}.

For the ``dark spot'' scenario, on the other hand,
only spot centers 
away from the center 
of the transit belt yield
proper fits to the transit lightcurve:
\mbox{$\theta \le 72\degr$} and \mbox{$\theta \ge 78\degr$}.
This is illustrated by the lower
panel of Fig.~\ref{fig:X2contours}
which also shows that
the resulting spot radii are smaller than
in the ``bright spot'' case 
(\mbox{$r = 4.6\degr \ldots 7.8\degr$}).

Also spots with centers outside the transit band yield
feasible solutions. 
An example is illustrated by the red arc in
Fig.~\ref{fig:TransitMap}, showing
solution DEQ of Table~\ref{tab:SpotSolutions}.
As illustrated by the figure, concerning area and longitude extent
they do not differ significantly from solutions with centers 
inside the band.
We do not discuss them further since
their radii do not describe the 
transited extension and area of the spot appropriately.


\begin{table}[t!]
  \begin{minipage}[h]{0.45\textwidth}
  \renewcommand{\footnoterule}{}
  \caption{
Parameters of characteristic spot solutions 
discussed in the text,
longitudes and colatitudes are given for the spot center.
          }        
    \label{tab:SpotSolutions}
\begin{center}
    \begin{tabular}{l c c c c c c}
      \hline \hline
       Model\footnote{
BC, BN and BS stand for 'bright central', 'bright north' and 'bright south', respectively.
They describe spots with colatitudes close to 
the center of the planetary disk.
DN and DS stand for 'north' and 'south' dark spot solutions, respectively;
DEQ represents a ``dark'' spot centered below the equator.
}
       & Long. 
       & Colat.
       & Radius
       & Flux \footnote{Relative to the photosphere}
       & $\chi^2$
       & Area \footnote{Fraction of total stellar surface} \\
      \hline
    BC & $216.4\degr$  &  $75.0\degr$  &  $7.8\degr$  & $0.75$  &  0.8 & 0.45\%  \\  
    BN & $216.5\degr$  &  $70.0\degr$  &  $9.5\degr$  & $0.75$  &  1.4 & 0.55\%  \\
    BS & $216.2\degr$  &  $80.0\degr$  &  $8.5\degr$  & $0.75$  &  1.2 & 0.47\%  \\
    DN  & $216.7\degr$  &  $71.0\degr$  &  $4.8\degr$  & $0.3$  &  1.7 & 0.18\%  \\
    DS  & $216.2\degr$  &  $81.0\degr$  &  $4.8\degr$  & $0.3$  &  1.9 & 0.18\%  \\
    DEQ  & $216.3\degr$  &  $94.0\degr$  &  $15.3\degr$  & $0.3$  &  1.6 & 0.72\%  \\
      \hline
      \hline
    \end{tabular}
\end{center}
  \end{minipage}
\end{table}

\section{Discussion}
\label{sec:Disc}
\vspace*{-0.85\medskipamount}
The shape of the transit lightcurves of CoRoT-2 exhibits 
highly significant
variations between different transits. This indicates a ubiquitous presence of 
starspots in the surface regions transited by the planet.
Furthermore, as illustrated by Fig.~\ref{fig:LCoverview} and the analysis of
\citet{Lanza09}  
the overall lightcurve of CoRoT-2 
continuously changes in amplitude
during the complete CoRoT observations.
This shows that the stellar surface of CoRoT-2a
persistently evolves on timescales shorter than its rotation period.
Such a fast spot evolution is interesting physically and makes 
CoRoT-2a a favourable case for the study of stellar activity, 
potentially a landmark system.

We concentrate our analysis on a single planetary transit whose lightcurve
shows a pronounced and isolated bump close to the transit center.
Assuming this bump is caused by a circular starspot,
we determine parameter ranges for this spot which 
reproduce the observed transit lightcurve.

While,
similar to \citeauthor{Pont07}'s analysis of HD189733,
the spot contrast is only weakly constrained, the spot longitude and
radius are closely confined by the transit lightcurve. 
The spot thus reconstructed on CoRoT-2b is comparable in
extent with
large spot groups on the Sun which cover up to about 1\% of the
solar surface
\citep{Baumann05, Norman05}.

Given the nearly $90\degr$ inclination of CoRoT-2b's orbit,
the transit-covered belt on its surface lies close to the equator.
Our analysis proves that CoRoT-2a, like the Sun, exhibits spots in this region.
Such well-constrained latitude measurements
near the equator are
difficult or impossible with other surface reconstruction
methods like Doppler imaging. Using long-term transit observations, this may 
allow to study activity cycles analog to the solar butterfly diagram.
Also, concerning possible indications of differential rotation on CoRoT-2a
\citep{Lanza09}, the spots in the transit-covered belt
could supply additional information.

Solar umbrae have diameters up to about 10~Mm, corresponding to one degree in 
heliographic coordinates; penumbrae reach approximately twice this size. 
Our study indicates that a surface resolution of potentially better 
than one degree can be achieved 
for host stars of eclipsing planets when applying transit mapping to
low-noise and fast-sampled lightcurves.
Thus, such observations offer a new 
opportunity to
study ``solar-like'' surface structures on other stars, they may 
for example allow to measure the umbra/penumbra contrasts
of their spots.


\vspace*{-\smallskipamount}
\begin{acknowledgements}
      U.W. acknowledges financial support from DLR, project 50 OR 0105.
\end{acknowledgements}

\end{document}